\documentclass[preprint,12pt]{elsarticle}

\usepackage{amsmath}    
\usepackage{graphicx}   
\usepackage{verbatim}   
\usepackage{color}      
\usepackage{hyperref}   
\usepackage{subfigure}

\journal{Nuclear Inst. and Methods in Physics Research, A}

\begin{document}

\begin{frontmatter}

\title{Characteristics of Signals Originating Near the Lithium-Diffused N+ Contact of High Purity Germanium P-Type Point Contact Detectors}


\author[pnnl]{E.~Aguayo}
\author[lbnleng]{M.~Amman}
\author[usc,ornl]{F.T.~Avignone~III}
\author[ITEP]{A.S.~Barabash}
\author[lbnl]{P.J.~Barton}		
\author[ornl]{J.R.~Beene}	
\author[ornl]{F.E.~Bertrand}
\author[lanl]{M.~Boswell}
\author[JINR]{V.~Brudanin}
\author[duke,tunl]{M.~Busch}
\author[lbnl]{Y-D.~Chan}
\author[sdsmt]{C.D.~Christofferson}
\author[uchic]{J.I.~Collar}
\author[ncsu,tunl]{D.C.~Combs}
\author[ornl]{R.J.~Cooper}
\author[lbnl]{J.A.~Detwiler}
\author[uw]{P.J.~Doe}
\author[ut]{Yu.~Efremenko}
\author[JINR]{V.~Egorov}
\author[ou]{H.~Ejiri}
\author[lanl]{S.R.~Elliott}
\author[duke,tunl]{J.~Esterline}
\author[pnnl]{J.E.~Fast}
\author[uchic]{N.~Fields}
\author[unc,tunl]{P.~Finnerty}
\author[unc,tunl]{F.M.~Fraenkle}
\author[ornl]{A.~Galindo-Uribarri}
\author[lanl]{V.M.~Gehman}
\author[unc,tunl]{G.K.~Giovanetti}
\author[unc,tunl]{M.P.~Green}
\author[usd]{V.E.~Guiseppe}
\author[JINR]{K.~Gusey}
\author[alberta]{A.L.~Hallin}
\author[ou]{R.~Hazama}
\author[unc,tunl]{R.~Henning}
\author[pnnl]{E.W.~Hoppe}
\author[sdsmt]{M.~Horton} 
\author[sdsmt]{S.~Howard}
\author[unc,tunl]{M.A.~Howe}
\author[uw]{R.A.~Johnson}
\author[blhill]{K.J.~Keeter}
\author[lanl]{M.F.~Kidd}
\author[uw]{A.~Knecht}
\author[JINR]{O.~Kochetov}
\author[ITEP]{S.I.~Konovalov}
\author[pnnl]{R.T.~Kouzes}
\author[pnnl]{B.D.~LaFerriere} 
\author[uw]{J.~Leon}
\author[ncsu,tunl]{L.E.~Leviner}
\author[lbnl]{J.C.~Loach}
\author[lbnl]{Q.~Looker}	
\author[lbnleng]{P.N.~Luke}
\author[unc,tunl]{S.~MacMullin}
\author[uw]{M.G.~Marino}
\author[lbnl]{R.D.~Martin \corref{cor1}}
\author[pnnl]{J.H.~Merriman}
\author[uw]{M.L.~Miller}
\author[usc,pnnl]{L.~Mizouni} 
\author[ou]{M.~Nomachi}
\author[pnnl]{J.L.~Orrell}
\author[pnnl]{N.R.~Overman}  
\author[usd]{G.~Perumpilly} 
\author[unc,tunl]{D.G.~Phillips II}
\author[lbnl]{A.W.P.~Poon}
\author[ornl]{D.C.~Radford}
\author[lanl]{K.~Rielage}
\author[uw]{R.G.H.~Robertson}
\author[lanl]{M.C.~Ronquest}
\author[uw]{A.G.~Schubert}
\author[ou]{T.~Shima}
\author[JINR]{M.~Shirchenko}
\author[unc,tunl]{K.J.~Snavely}
\author[lanl]{D.~Steele}
\author[unc,tunl]{J.~Strain}
\author[JINR]{V.~Timkin}
\author[duke,tunl]{W.~Tornow}
\author[ornl]{R.L.~Varner}
\author[lbnl]{K.~Vetter\fnref{ucb}}
\author[unc,tunl]{K.~Vorren}
\author[unc,tunl,ornl]{J.F.~Wilkerson}  
\author[JINR]{E.~Yakushev}
\author[lbnleng]{H.~Yaver}	
\author[ncsu,tunl]{A.R.~Young}
\author[ornl]{C.-H.~Yu}
\author[ITEP]{V.~Yumatov}
\author{\newline (The {\sc Majorana} Collaboration)}

\cortext[cor1]{Corresponding author, rdmartin@lbl.gov, phone: +1 510 486 7121, fax: +1 510 486 6738}
\address[pnnl]{Pacific Northwest National Laboratory, Richland, WA, USA}
\address[lbnleng]{Engineering Division, Lawrence Berkeley National Laboratory, Berkeley, CA, USA}
\address[usc]{Department of Physics and Astronomy, University of South Carolina, Columbia, SC, USA}
\address[ornl]{Oak Ridge National Laboratory, Oak Ridge, TN, USA}
\address[ITEP]{Institute for Theoretical and Experimental Physics, Moscow, Russia}
\address[lbnl]{Nuclear Science Division, Lawrence Berkeley National Laboratory, Berkeley, CA, USA}
\address[lanl]{Los Alamos National Laboratory, Los Alamos, NM, USA}
\address[JINR]{Joint Institute for Nuclear Research, Dubna, Russia}
\address[duke]{Department of Physics, Duke University, Durham, NC, USA}
\address[tunl]{Triangle Universities Nuclear Laboratory, Durham, NC, USA}
\address[sdsmt]{South Dakota School of Mines and Technology, Rapid City, SD, USA}
\address[uchic]{Department of Physics, University of Chicago, Chicago, IL, USA}
\address[ncsu]{Department of Physics, North Carolina State University, Raleigh, NC, USA}
\address[uw]{Center for Experimental Nuclear Physics and Astrophysics, and Department of Physics, University of Washington, Seattle, WA, USA}
\address[ut]{Department of Physics and Astronomy, University of Tennessee, Knoxville, TN, USA}
\address[ou]{Research Center for Nuclear Physics and Department of Physics, Osaka University, Ibaraki, Osaka, Japan}
\address[unc]{Department of Physics and Astronomy, University of North Carolina, Chapel Hill, NC, USA}
\address[usd]{Department of Physics, University of South Dakota, Vermillion, SD, USA} 
\address[alberta]{Centre for Particle Physics, University of Alberta, Edmonton, AB, Canada}
\fntext[ucb]{Alternate Address: Department of Nuclear Engineering, University of California, Berkeley, CA, USA}
\address[blhill]{Department of Physics, Black Hills State University, Spearfish, SD, USA}


\date{\today}

\begin{abstract}

A study of signals originating near the lithium-diffused n+ contact of p-type point contact (PPC) high purity germanium detectors (HPGe) is presented. The transition region between the active germanium and the fully dead layer of the n+ contact is examined. Energy depositions in this transition region are shown to result in partial charge collection. This provides a mechanism for events with a well defined energy to contribute to the continuum of the energy spectrum at lower energies. A novel technique to quantify the contribution from this source of background is introduced. Experiments that operate germanium detectors with a very low energy threshold may benefit from the methods presented herein.

\end{abstract}


\end{frontmatter}

\section{Introduction}
P-type point contact (PPC) high purity germanium (HPGe) detectors \cite{luke_89, Barbeau:2007qi} are of interest for use in astroparticle physics. The small area of the point contact (see Fig. \ref{fig:DUPPC_Crystal} for a diagram) results in a sharply peaked weighting potential near the point contact as well as a small readout capacitance. The localized weighting potential of PPC detectors results in distinct current pulses from individual interaction charge clouds, and enables pulse-shape analysis to be used to distinguish between single-site and multiple-site events. Backgrounds from gamma rays scattering multiple times in a detector can thus be identified \cite{Barbeau:2007qi,Budjas:2009zu,Cooper2011303} resulting in this technology being adopted by the {\sc Majorana} \cite{Schubert:2011nm, Aguayo:2011sr, MAJORANA:2011aa,  MJGeneral:2010i} and GERDA \cite{Agostini:2010ke} experiments to search for the neutrinoless double-beta decay of $^{76}$Ge. 

The small readout capacitance allows PPC detectors to be operated with low levels of electronic noise and a correspondingly low energy threshold. This makes them attractive candidates for light dark matter searches, such as the CoGeNT experiment \cite{Barbeau:2007qi, Aalseth:2010vx} which has demonstrated the operation of a PPC detector with a threshold of $\sim 400$\,eV. The low energy threshold of these detectors has also made them candidates for detecting coherent neutrino-nucleus scattering \cite{Barbeau:2007qi, Wong:2008vk, Wong:2010zzc}. Searches for neutrinoless double-beta decay using $^{76}$Ge also benefit from the low energy threshold of PPC detectors; x-rays at 1.3\,keV and 10.4\,keV following the electron-capture decay of $^{68}$Ge can be used to tag the subsequent $\beta$+ decay of $^{68}$Ga which has a Q-value of 2.9\,MeV, above the region of interest for neutrinoless double-beta decay. Understanding the behavior of these detectors at low energies is critical for all these types of experiments.

A recent publication by the CoGeNT collaboration \cite{Aalseth:2010vx} presented results from a PPC detector used to examine x-rays with energies of 10.4\,keV from the decays of internal $^{71}$Ge (produced by thermal neutron activation of the detector using an americium-beryllium source). X-rays that deposited energy near the thick, lithium-diffused, n+ outer contact of their detector were shown to result in signals with incomplete charge collection at energies between 2.5\,keV and 7.5\,keV (Fig. 1 in Ref. \cite{Aalseth:2010vx}). Their observation is consistent with the existence of a transition layer, between the bulk active germanium and the contact dead layer, where a weak electric field results in slow and incomplete charge collection. Signals originating in this transition layer give rise to characteristically slow charge pulses and contribute to the continuum at lower energies. Understanding the properties of energy depositions in the transition region is important for determining the active volume of a detector as well as for understanding how the incomplete charge collection of these events contributes to the continuum in the energy spectrum.

The existence of the transition layer between the contact and the active germanium has been established in the literature, although a comprehensive picture is lacking and past research has been somewhat disconnected. Pulses resulting from anomalous charge collection were identified in the early development of lithium-drifted germanium Ge(Li) detectors. Soon after the initial characterization \cite{tavendale:1963, ewan:1964} of Ge(Li) detectors in 1963 by Ewan and Tavendale, Alexander \textit{et al.} \cite{alexander:1964} observed a class of events with abnormally long collection times, and attributed these to interactions occurring between the fully compensated (active) part of their detector and the thick (dead) n+ contact, where the electric field is too weak to collect charge carriers efficiently. They warned that any electron interactions near the n+ contact, including interactions from electrons produced when high energy gamma rays interact in the surroundings of the detector, would contribute to the continuum of the energy spectrum.

In 1967, Tamm \textit{et al.} \cite{tamm:1967} identified diffusion as the principal mechanism for charge carriers created in the transition layer to enter the region where the electric field is strong enough for effective charge collection. The diffusion coefficient, the depth of the un-depleted region, and the charge carrier lifetime were used to estimate the diffusion time to be of order microseconds, giving rise to slow pulses and giving enough time for charge loss due to recombination. Tamm \textit{et al.} showed that the background from slow pulses increased at lower energies in their energy spectra. Detailed experimental studies by Strauss and Larsen \cite{strauss:1967} in the same year and by Sakai \cite{sakai:1971} in 1971 confirmed, using a variety of detector geometries and collimated sources, that the slow pulses originated near the contacts.

More recent studies of germanium detectors, focused on efficiency measurements and the detailed modeling of energy spectra, have also examined the properties of events near the lithium-diffused contacts. In 1989,  Clouvas \textit{et al.} \cite{clouvas:1998} noted that the efficiency measurements for their detector yielded a dead layer thickness of 2.5\,mm that did not agree with the value of 0.5\,mm provided by the manufacturer. They additionally found that in order to model the energy spectrum from their p-type coaxial detector at low energies, they needed to include a transition layer where partial charge collection occurred. By considering the diffusion of the lithium in the n+ contact, they derived a model for the size of the transition region that enabled them to successfully simulate their energy spectrum at lower energies.

In 2003, Rodenas \textit{et al.} \cite{rodenas:2003} showed that the lithium-diffused layer changes both the active volume of the detector, and the efficiency for detecting low energy radiation due to the increased attenuation from the dead layer. They pointed out that the thickness of this layer is difficult to measure because of the partially active transition region and that the values provided by the manufacturers are often wrong. In 2005, Maleka \textit{et al.} \cite{maleka:2005} used Monte Carlo simulations to show that small changes in the thickness of the lithium-diffused layer can result in significant changes in the efficiency for low energy gamma rays. Dryak \textit{et al.} \cite{dryak:2010} argued that neglecting the transition layer results in underestimates of the total efficiency of a detector in their work from 2010. They also showed evidence, using a coincidence measurement, that signals from the transition layer were delayed in time compared to normal signals, which is consistent with diffusion being the dominant transport mechanism for charge carriers in the transition layer. The {\sc Majorana} collaboration is performing detailed measurements using coincidences between betas and gammas and between gammas to study the precise mechanism of charge carrier transport in this transition layer using the detector described in \cite{Aalseth2011692} and another similar detector produced by CANBERRA. Results of these studies will be presented in a subsequent publication.

  Studies between 2007 and 2011 by Huy \textit{et al.} \cite{huy:2007} and by Huy \cite{huy:2010,huy:2011} showed that the depth of the lithium-diffused layer increased as their HPGe detector aged, due to diffusion of the lithium atoms over time. They noted that this effect is particularly pronounced for periods where the detector was kept at room temperature. Using these data, they showed that the counts in the lower part of their energy spectrum increased with the depth of the lithium-diffused layer. In 2010, Luis \textit{et al.} \cite{luis:2010} observed significant discrepancies between the dead layer they measured on a Broad Energy Germanium (BEGe) detector and that quoted by the manufacturer (CANBERRA \cite{CanberraWeb}); they attributed the discrepancy to an increase in dead layer thickness during the time between their measurement and the one performed by CANBERRA.

In summary, there is substantial evidence in the literature to support the claim that high energy events can contaminate the lower part of the energy spectrum continuum in germanium detectors that have a thick, lithium-diffused, contact. These events occur in a transition region between the fully active germanium and the dead outer contact. Pulse shapes from these events are characteristically slower than those originating in the active region, delayed in time and have a degraded pulse height. These observations are consistent with the hypotheses that the dominant transport mechanism for charge carriers created in the transition region is diffusion into the bulk region, followed by drift in the electric field. Furthermore, the thickness of the lithium-diffused region can change in time due to continuous diffusion of lithium.

This paper presents the results of a study that characterized events originating near the lithium-diffused contact of
a custom-made PPC detector illuminated with a collimated source of low energy gamma rays. The existence of a transition region between the fully active germanium and fully dead layer of the n+ lithium-diffused contact is verified. Energy depositions in this region are observed to result in pulses with slow rise times and degraded charge collection. A novel method for determining the depth of the transition layer and the depth at which energy depositions become fully detectable is introduced. The method is also applied to the detector described in \cite{Aalseth2011692} and it is shown that the size of the transition and dead layers is dependent on the process that was used to fabricate the lithium-diffused contact.

\section{Description of the LBNL PPC detector and data acquisition}
\begin{figure}[!htbp]
\centering
\includegraphics[width=0.8\textwidth]{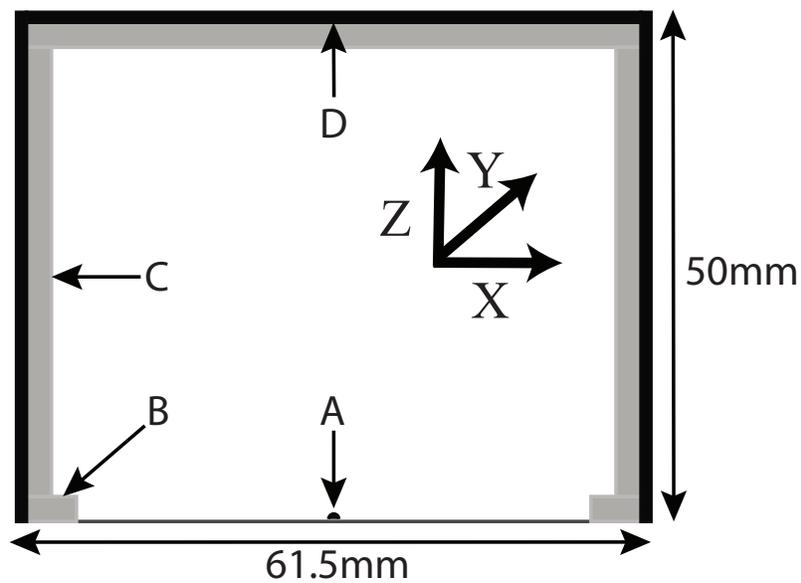} 
\caption{\label{fig:DUPPC_Crystal}Schematic diagram of the LBNL PPC crystal, showing the point contact (A), the 5\,mm lithium-diffused ``wraparound'' (B), the lithium-diffused layer (C), and the aluminized low resistance contact for the high-voltage (D). The coordinate system referred to in the text is also shown, with the y axis going into the page and the origin at the point contact. The aspect ratio of the detector is correct, although the size of the point contact (1\,mm diameter), and the thickness of the lithium and metalization layers are not to scale.}

\end{figure}

A custom PPC detector \cite{DriftTime_paper_2011}, illustrated in Fig. \ref{fig:DUPPC_Crystal}, was fabricated at Lawrence Berkeley National Laboratory from a high purity germanium crystal purchased from ORTEC \cite{OrtecWeb}. The crystal has a radius of 31\,mm and a height of 50\,mm. A lithium-diffused layer covers the outside of the entire crystal except for the surface with the point contact, where it only covers the outer 5\,mm (the ``lithium wraparound''). The lithium-diffused surface (except for the wraparound) was metalized by vapor deposition of aluminum to create a low resistance high voltage outer contact. A hemispherical dimple with a diameter of 1.0 $\pm$ 0.5\,mm was machined into the crystal, sputter coated with amorphous silicon \cite{amman:2007}, and metalized by vapor deposition of aluminum to define the point contact. The detector was operated with an un-passivated surface between the point contact and the lithium-diffused wraparound.

The crystal was installed in a custom copper mount prototype designed for the {\sc Majorana Demonstrator} experiment \cite{Schubert:2011nm, Aguayo:2011sr, MAJORANA:2011aa,  MJGeneral:2010i}. The mount was fixed to a copper cold plate and surrounded by a 1.5\,mm thick copper infrared shield to keep the detector cold. The detector assembly was housed in an aluminum cryostat with a wall thickness of 2.2\,mm and cooled with liquid nitrogen. Charge pulses from the point contact were processed with a cold FET mounted on a low mass board designed for low-noise performance and low radioactivity levels \cite{LMFE_paper_2011}. The amplification loop was closed outside the cryostat in a custom charge sensitive resistive-feedback preamplifier with a 16\,MHz bandwidth. Charge pulses processed by the preamplifier were digitized using a Struck SIS3302 16-bit 100\,MHz digitizing card. 

The detector was scanned with 59.5\,keV gamma rays from a 1\,mCi $^{241}$Am calibration source collimated using tungsten collimators (10\,mm thick, hole diameters between 0.5\,mm and 2\,mm). Two scans were performed; one in the $xy$ plane (illuminating the point contact surface of the detector) and one in the $yz$ plane (illuminating the side of the detector). Readers are referred to Fig. \ref{fig:DUPPC_Crystal} for the definition of the coordinate system used in this paper. Two different setups (collimators, source positioning system) were used for the two scans. Consequently, the beam spot sizes on the crystal were different in the two scans. The diameter of the beam was determined from geometry to be 3\,mm $\pm$ 1\,mm in the $xy$ scan and 2.0\,mm $\pm$ 1,0\,mm  in the $yz$ scan. In each scan, data were taken at various positions of the collimated source for a fixed duration of 4 minutes.

\section{Determination of the full charge collection depth}
\label{sec:lithickness}

 The thickness of the lithium-diffused layer of a detector is difficult to know accurately from the manufacturing process. The n+ contact on the LBNL PPC detector was fabricated by first evaporating lithium onto the surface of the crystal (masking the areas where no lithium was desired). The crystal was then heated up to approximately 280$^\circ$C in order to increase the diffusion of lithium atoms into the crystal. Variables such as oxygen concentration at the surface, and the temperature profile of the crystal as a function of time during the diffusion process result in large uncertainties when trying to determine the lithium thickness from the manufacturing parameters.
 
A technique is presented here to characterize the thickness of the region where the lithium diffusion affects energy collection, usually referred to as the ``dead layer''. The method is similar to that presented in \cite{boson:2008} and \cite{dryak:2006} in which a collimated beam of gamma rays was incident on the detector surface at different angles.

It will be apparent from this work that the term ``dead layer'' must be defined appropriately, and that there are three principal regions that must be distinguished. The outermost region of the contact, which is conductive and has no electric field, may be considered ``dead''. There is then a transition region where energy depositions result in incomplete charge collection. The transition layer depth (TLD) is defined as the depth at which this region begins and corresponds to the boundary with the dead layer. Finally, the full charge collection depth (FCCD) is defined as the depth at which the truly active part of the germanium detector begins and corresponds to the boundary of the transition region and the region where full charge collection occurs. Higher energy depositions can result in charge carrier clouds that extend across the different regions. In this case, the different parts of the charge cloud will be collected differently, depending on the region in which they were created. The 59.5\,keV gamma rays used in this work result in electrons with a range of less than 20\,$\mu$m in germanium, and the associated charge carrier clouds can effectively be taken as point-like.

\begin{figure}[!htbp]
\centering
\includegraphics[width=0.9\textwidth]{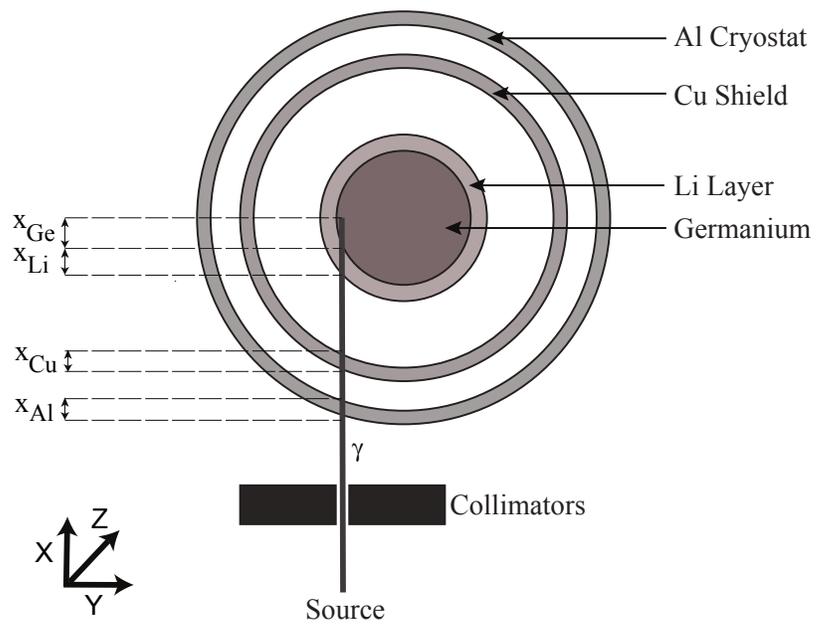} 
\caption{\label{fig:geometry_yz_full} Projection of the setup used for the scan of the side of the detector, seen from below ($z$-axis into the page). The collimated beam of 59.5\,keV gamma rays in the $x$ direction is shown traversing the aluminum cryostat, the copper infrared shield, the lithium-diffused layer and the germanium. As the $y$-position of the collimator and source are varied, the path length of the beam through the different attenuating layers changes.}
\end{figure}

Fig. \ref{fig:geometry_yz_full} shows a bottom-view of the setup that was used to scan the side of the detector in this work. The beam from the source was collimated in the x-direction. The copper infrared shield and the aluminum cryostat are also shown. It can be seen that, as the $y$ position of the source is varied, the path length of the beam through the lithium-diffused layer, $x_{\mathrm{Li}}(y)$, changes. Given the attenuation coefficients \cite{NISTWeb} for copper, $\mu_{\mathrm{Cu}}=1.63$\,cm$^2$/g, aluminum, $\mu_{\mathrm{Al}}=0.281$\,cm$^2$/g, and germanium, $\mu_{\mathrm{Ge}}=2.07$\,cm$^2$/g, the beam attenuation from the various materials can be determined. The attenuation coefficient of the lithium-diffused layer is taken as equal to that of germanium.  A simple model for the number of counts in the full energy peak as a function of the $y$ position of the beam, $N(y)$, is given by:

\begin{equation}
\label{eqn:countsVsY}
N(y) = N_0 \exp\left[-\left(\mu_{\mathrm{Al}}\rho_{\mathrm{Al}}x_{\mathrm{Al}}(y)+\mu_{\mathrm{Cu}}\rho_{\mathrm{Cu}}x_{\mathrm{Cu}}(y)+\mu_{\mathrm{Ge}}\rho_{\mathrm{Ge}}x_{\mathrm{Li}}(y)\right)\right]
\end{equation}
where $N_0$ is the number of full energy counts that would be observed in the detector without attenuation, $\rho_\mathrm{X}$ is the density of material X and $x_{\mathrm{X}}(y)$ is the path length of the beam through material X and depends on the position of the beam, $y$. By fitting $N(y)$ to data, $x_{\mathrm{Li}}(y=0)$ can be determined and identified with the minimum depth at which the full energy of the gamma rays is detected. This corresponds to the FCCD (full charge collection depth, defined above) which is traditionally identified as the dead layer thickness and determined by efficiency measurements.

\begin{figure}[!htbp]
\centering
\includegraphics[width=0.8\textwidth]{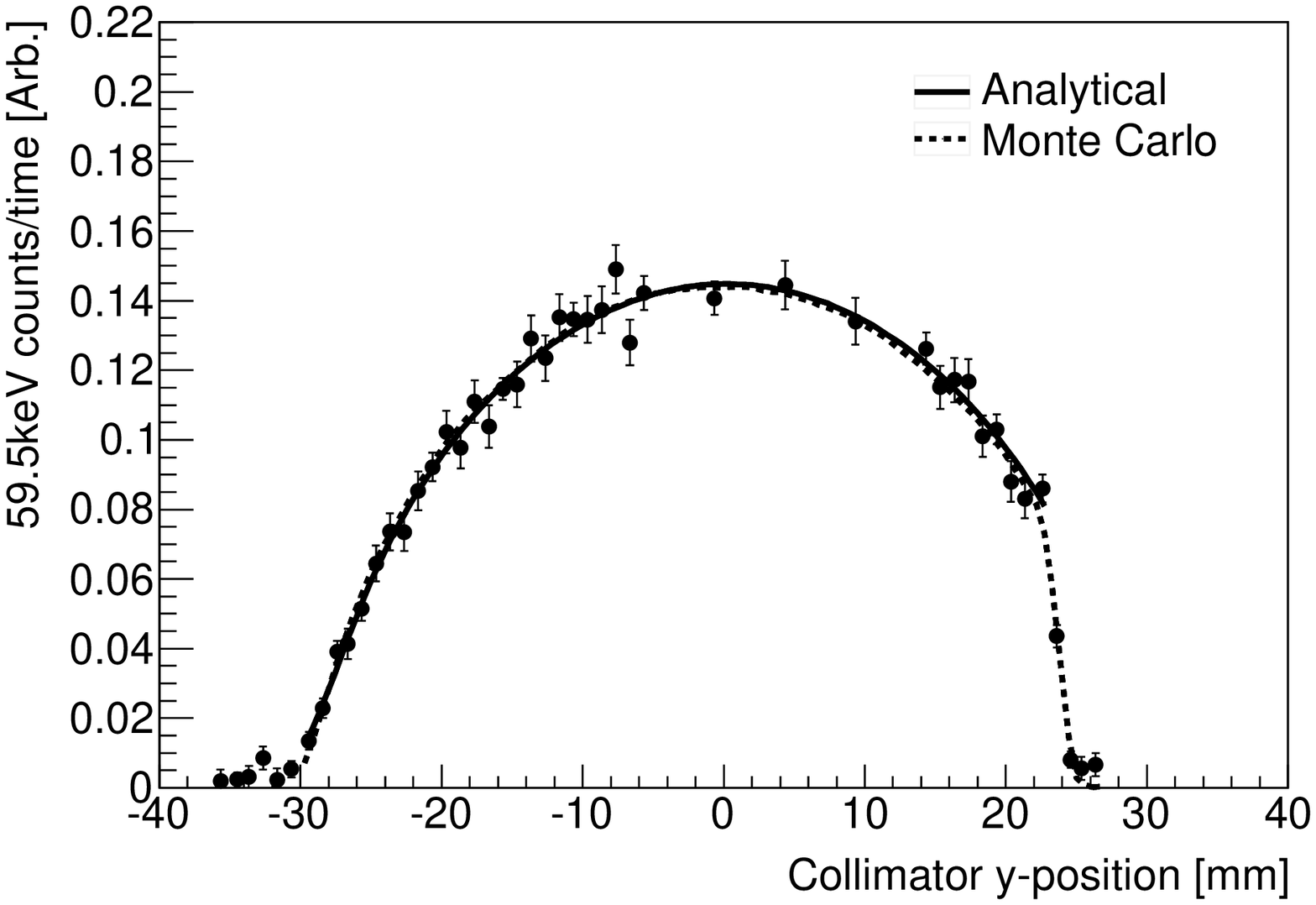}
\caption{\label{fig:LiThicknessFit} The rate of events in the 59.5\,keV photopeak as a function of the $y$-position of the collimated $^{241}$Am source, from the setup illustrated in Fig. \ref{fig:geometry_yz_full}. The solid line shows a fit to the data using the the analytical function in equation \ref{eqn:countsVsY} where the intensity of the beam, the center of the crystal, the thickness of the germanium dead layer, and the angle between the collimator and crystal y-axes were allowed to vary. The data were not fit beyond $y=20$\,mm, as the analytical model does not take into account the sharp fall off in events caused by a copper rod in the detector mount. The depth of the dead plus transition layer from the fit to the analytical function was determined to be 0.77\,mm $\pm$ 0.12\,mm. The dashed line shows a fit of the same data to the probability distribution functions obtained from a Monte Carlo simulation of the setup. The Monte Carlo model included the copper rod in the detector mount and reproduced the data beyond $y=20$\,mm. The depth determined using the Monte Carlo was 0.78\,mm $\pm$ 0.08\,mm, in excellent agreement with the analytical fit.}

\end{figure}

Fig. \ref{fig:LiThicknessFit} shows data collected from a scan of the side of the LBNL PPC detector along the $y$ direction fit to the function in equation \ref{eqn:countsVsY} (solid line) to determine FCCD. At each position of the collimated source, the rate of events in the full energy photopeak at 59.5\,keV was determined by integrating the energy spectrum near the peak (between 58\,keV and 61\,keV) and subtracting a background obtained from counts in the region above the peak, between 61\,keV and 80\,keV to avoid any contribution from energy-degraded events in the peak. The unattenuated intensity, $N_0$, the position of the center of the crystal in relation to the collimator position, and FCCD were determined from a fit performed using the MINUIT utilities in the  ROOT software package from CERN \cite{ROOTTminuitWeb}. Parameters, such as the attenuation coefficients, densities and positions of other materials (aluminum, copper) were fixed to their known values. In order to account for a finite beam spot size, $N(y)$ was convolved with a Gaussian. The fit to the data produced the best agreement when the standard deviation of the gaussian was equal to 2\,mm. Systematic uncertainties from possible errors in parallax (to account for a possible misalignment of the beam with respect to the crystal coordinate system) were found to be the dominant systematic error in the determination of FCCD. To account for this, the angle between the y-axis of the collimator and the y-axis of the crystal was varied in the fit so that the uncertainty on FCCD from the fit includes this systematic contribution. The effect of changing the dimensions of the cryostat, the IR shield, and the germanium crystal was also examined but it was found that these dimensions are very well constrained by the fit and do not affect the uncertainty in determining FCCD. The full charge collection depth was determined to be 0.77\,mm $\pm$ 0.12\,mm. Since the dominant systematic parameters were also varied in the fit, the uncertainty includes systematic and statistical contributions. If the angle between the y-axes of the collimator and crystal was not allowed to vary, FCCD was determined to be 0.77\,mm $\pm$ 0.06\,mm, which is representative of the statistical uncertainty. The sharp decrease in the number of counts at  $y\approx 20$\,mm in Fig. \ref{fig:LiThicknessFit} is due to the presence of a copper rod that is part of the crystal mount and was not incorporated into this simple analytical model.

\begin{figure}[!htbp]
\centering
\includegraphics[width=0.8\textwidth]{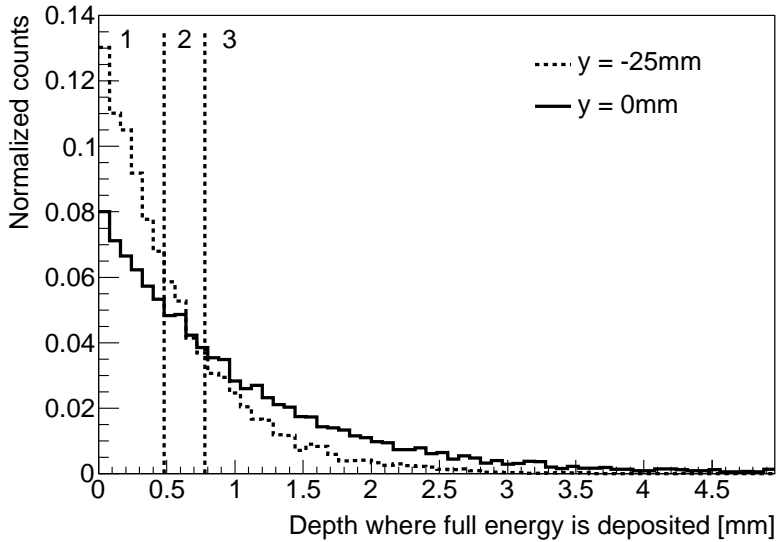}
\caption{\label{fig:MCPDFS} Normalized distributions of the depth at which Monte Carlo events have deposited their full energy for two different $y$-positions of the collimated source. The collimator position of $y=0$\,mm corresponds to the center of the crystal and the position of $y=-25$\,mm is closer to the edge (Fig. \ref{fig:geometry_yz_full}). The distributions show that by moving the collimated source, different depth profiles in the crystal are sampled. The area above 0.78\,mm (region 3) was determined by the fit in Fig. \ref{fig:LiThicknessFit} and corresponds to events where the full energy is detected. The region between 0.48\,mm and 0.78\,mm (region 2) was determined from the fits in Fig. \ref{fig:TransitionFitComp} and corresponds to a transition region where events have an energy degraded down to 26\,keV and slow rise times. The region below 0.48\,mm (region 1) corresponds to events where less than 26\,keV is detected and would correspond to a fully dead layer for a detector with a 26\,keV threshold.}

\end{figure}

\begin{figure}[!htbp]
\centering
\includegraphics[width=0.8\textwidth]{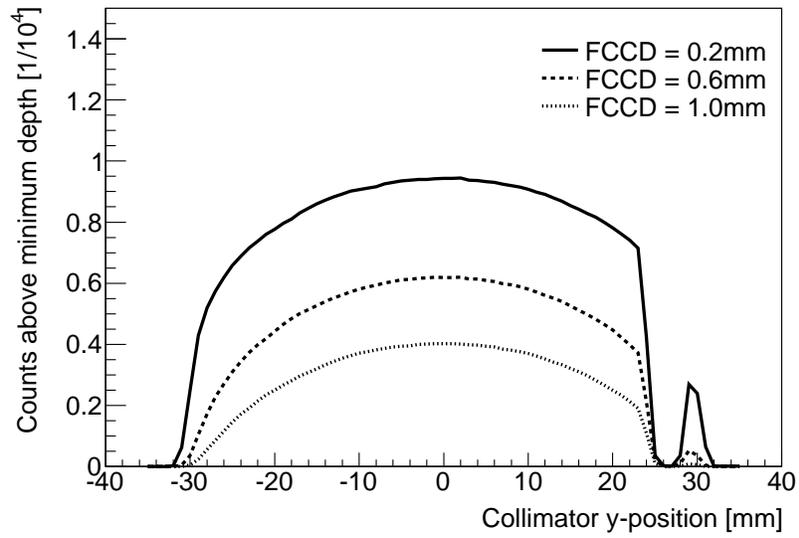}
\caption{\label{fig:mcdepthpdfscomp} Monte Carlo probability distribution functions (PDFs) of the number of full energy counts as a function of the collimator $y$-position for three different choices of FCCD. The PDFs were determined from the total number of counts above a given depth, FCCD, in the distributions shown in Fig. \ref{fig:MCPDFS} (corresponding to region 3 in Fig. \ref{fig:MCPDFS}). A function that interpolates between these PDFs was used to fit the data in Fig. \ref{fig:LiThicknessFit}.}

\end{figure}

In order to verify the fit to the analytical function of equation \ref{eqn:countsVsY} and confirm this model, a Monte Carlo simulation of the setup in Fig. \ref{fig:geometry_yz_full} was performed using Geant4 version 4.9.2 \cite{geant4}. The geometry was reproduced in the model and a collimated beam of 59.5\,keV gamma rays was simulated. The output of the Monte Carlo simulation was the distribution of energy depositions (hits) in the crystal. Monte Carlo runs were performed by generating a fixed number of 59.5\,keV gamma rays at each position of the collimator and source. For each position of the collimator, the distribution of the depth in the crystal at which events deposit their full energy was determined. This distribution is illustrated for two different $y$-positions (0\,mm and -25\,mm) of the collimator in Fig. \ref{fig:MCPDFS}. Different depth profiles in the detector can be sampled by placing the source and collimators at different positions. In order to model the full charge collection depth, only events that deposit their full energy at a depth greater than FCCD (region 3 in Fig. \ref{fig:MCPDFS}) are considered to contribute to the full energy photopeak in the data. The number of events in the distribution illustrated in Fig. \ref{fig:MCPDFS} above a given depth, FCCD, was used to build a probability distribution function (PDF) for the number of events in the photopeak as a function of the $y$-position of the collimator. The shapes of the PDFs are very sensitive to the value of FCCD. These PDFs were built for a range of values of the full charge collection depth, as illustrated in Fig. \ref{fig:mcdepthpdfscomp} for FCCD$=0.2$\,mm, FCCD$=0.6$\,mm and FCCD$=1.2$\,mm.

A function that numerically interpolates between the PDFs was constructed and used to fit the data in Fig. \ref{fig:LiThicknessFit} using the MINUIT utility in the ROOT software package. The resulting fit is shown by the dashed line in Fig. \ref{fig:LiThicknessFit}. The PDFs for the different depths were normalized to have the same number of counts at $y=0$ and the fit function included an overall normalization to the number of events in the data as well as a parameter to allow the position of the center of the crystal to vary. Because the copper rod in the detector mount was included in the simulation, the range of the fit was extended in the positive $y$-direction and reproduced the decrease in the number of counts from a copper rod in the detector mount at $y\approx 20$\,mm.

 Systematic uncertainties were examined by varying the size of the collimator in the Monte Carlo geometry as well as by tilting the beam below the $x$-axis and found to be negligible. Systematic effects from parallax were examined by letting the angle between the collimator $y$-axis and the detector $y$-axis vary in the fit and were found to be the dominant contribution to the systematic uncertainty. With systematic and statistical uncertainties included, FCCD was determined to be 0.78\,mm $\pm$ 0.08\,mm, in excellent agreement with the fit using the analytical function.

\section{Evidence that slow, energy degraded pulses arise near the lithium-diffused surfaces}

\begin{figure}[!htbp]
\centering
\includegraphics[width=0.8\textwidth]{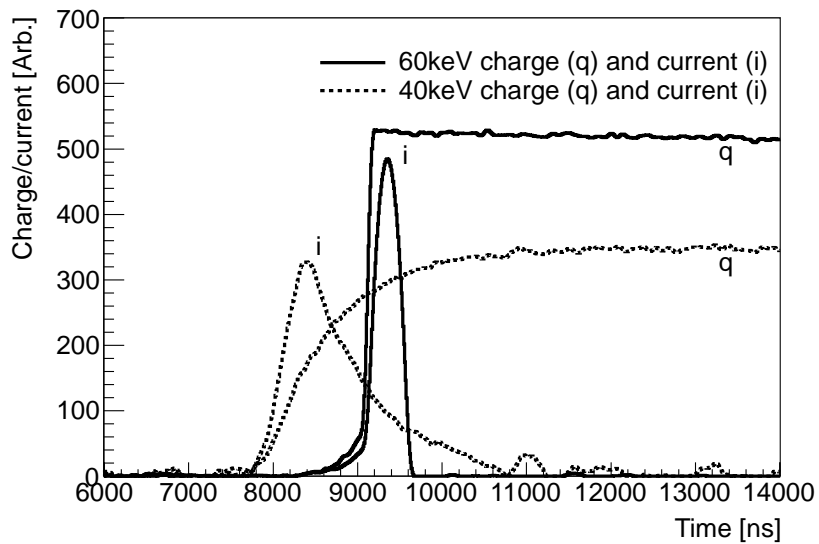} 
\caption{\label{fig:PulseComparison}Comparison of the charge and current pulses for a 60\,keV (solid) and an energy-degraded 40\,keV event (dashed). The current pulses were calculated from the charge pulses using the method described in \cite{DriftTime_paper_2011} with an averaging of 500\,ns. The lower energy event is a characteristic slow pulse that arises from the transition layer. It is easily selected based on either the rising time of the charge waveform or the asymmetry of the current pulse. The 10\% to 80\% rise time of the slow charge pulse is 1400\,ns compared to 130\,ns for the typical 60\,keV pulse illustrated here. For legibility in this figure, the waveforms were smoothed by a Gaussian filter.}

\end{figure}

Typical charge pulses from point contact detectors have a characteristic shape due to the sharply peaked weighting potential near the point contact. Pulses show an initial, slow, smaller amplitude component from the charge carriers drifting in a region of low weighting potential. This is followed by a larger, fast component from the short amount of time that the charge carriers spend in the region of high weighting potential near the point contact. A typical charge pulse from a 59.5\,keV event in the PPC detector used in this work is shown in Fig. \ref{fig:PulseComparison}, along with its corresponding current pulse calculated with a 500\,ns averaging time using the method described in \cite{DriftTime_paper_2011} (solid lines). Also shown in Fig. \ref{fig:PulseComparison} is a characteristically slow charge pulse along with its corresponding current pulse (dashed line). A simple metric to distinguish these types of events is either to measure the time that a charge waveform takes to rise from 10\% to 80\% or to use the asymmetry in the rise and fall times of the corresponding current pulse. 

 The shape of the slow pulse cannot be explained by the drift of a single cloud of charge carriers to the point contact from any location in the crystal, as this would produce a charge pulse with a fast rise time. Such a long rise time indicates that the charge carriers arrive at the point contact over a period of time.

\begin{figure}[!htbp]
\centering
\includegraphics[width=0.8\textwidth]{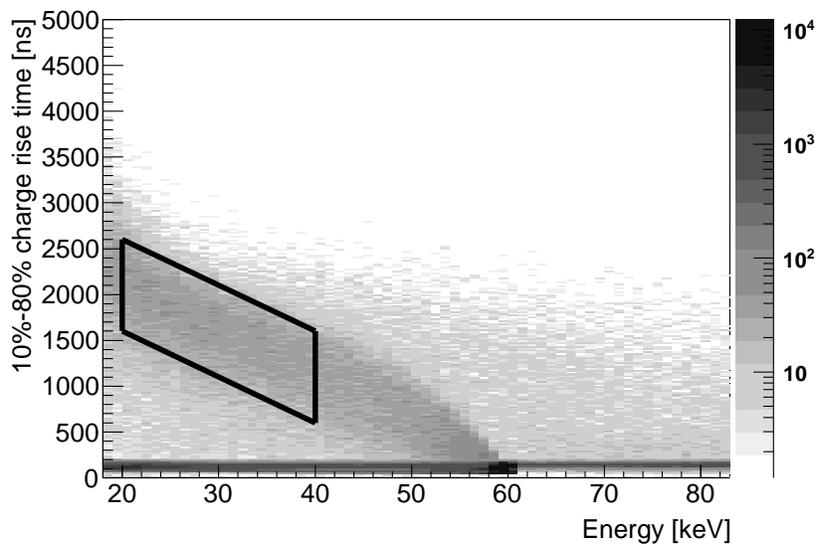} 
\caption{\label{fig:EnergyVsRiseTime_XY}Distribution of the 10\% to 80\% rise times of charge waveforms as a function of the measured energy in the $^{241}$Am source scan of the point contact face of the detector. A population of slow, energy degraded, (SED) pulses is evident as the diagonal band emerging from the 59.5\,keV photopeak. A box is shown to select a clean sample of these events; most of the data (83\%) lie in the band below 300\,ns.}

\end{figure}

The data collected by scanning the point contact surface with the collimated $^{241}$Am source can be used to confirm the origin of the slow pulses. Fig. \ref{fig:EnergyVsRiseTime_XY} shows a histogram of the distribution of the 10\% to 80\% charge waveform rise times as a function of the measured energy of the events in this scan. This is similar to Fig. 2 in Ref. \cite{Aalseth:2010vx}. The distribution shows a diagonal band of events originating from the 59.5\,keV $^{241}$Am peak with degraded energy and longer rise times. A box is drawn on the figure to highlight a clean sample of these events.

\begin{figure}[!htbp]
\centering
\includegraphics[width=0.8\textwidth]{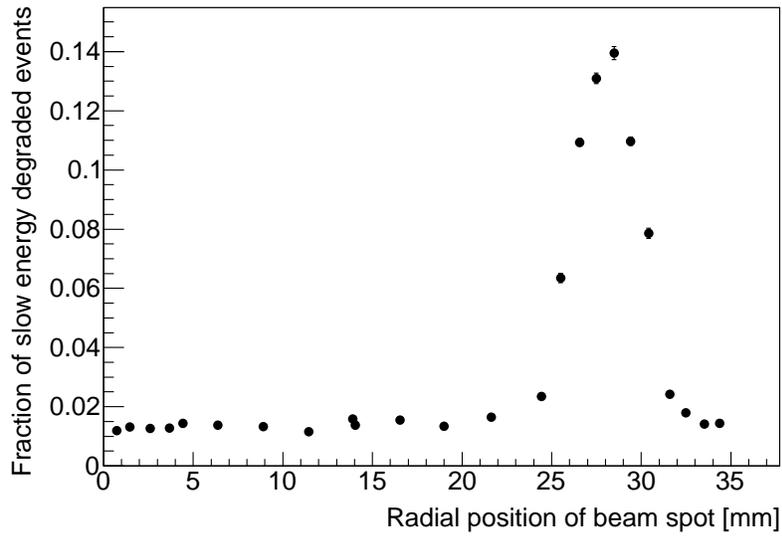} 
\caption{\label{fig:SlowPulse1D_Y_XY}The fraction of SED pulses in the scan of the point contact surface of the detector as a function of the radial position of the collimated beam (the beam points upwards when referring to Fig. \ref{fig:DUPPC_Crystal} and a radius of zero corresponds to the point contact). For a given position of the collimator, the fraction of slow pulses is taken as the number of events in the box shown in Fig. \ref{fig:EnergyVsRiseTime_XY} divided by the number of events in the full energy photopeak at 59.5\,keV. The fraction of SED pulses increases dramatically at the edge of the crystal, near a radius of 25\,mm, where the 5\,mm-wide lithium wraparound is located, confirming that these are related to charge depositions near the lithium-diffused region of germanium. The diameter of the beam spot was approximately 3\,mm in these data. }

\end{figure}

Fig. \ref{fig:SlowPulse1D_Y_XY} shows the fraction of events in this box relative to the number of events in the 59.5\,keV photopeak as a function of the radial position of the collimated beam on the surface of the detector. The fraction of slow, energy degraded (SED) pulses rises dramatically near the edges of the crystal. This rise in the fraction of SED events occurs at the outer 5\,mm, where the lithium-diffused wraparound is located (Fig. \ref{fig:DUPPC_Crystal}). This result shows that significantly more SED pulses arise in the regions of a detector where a thick lithium-diffused contact is present. This is consistent with the interpretation in Ref. \cite{Aalseth:2010vx} that there exists a transition region in the lithium-diffused layer where partial charge collection occurs (degrading the energy), and that pulse shapes are slower as the charge carriers take time to diffuse into the fully active region of the crystal.

\begin{figure}[!htbp]
\centering
\includegraphics[width=0.8\textwidth]{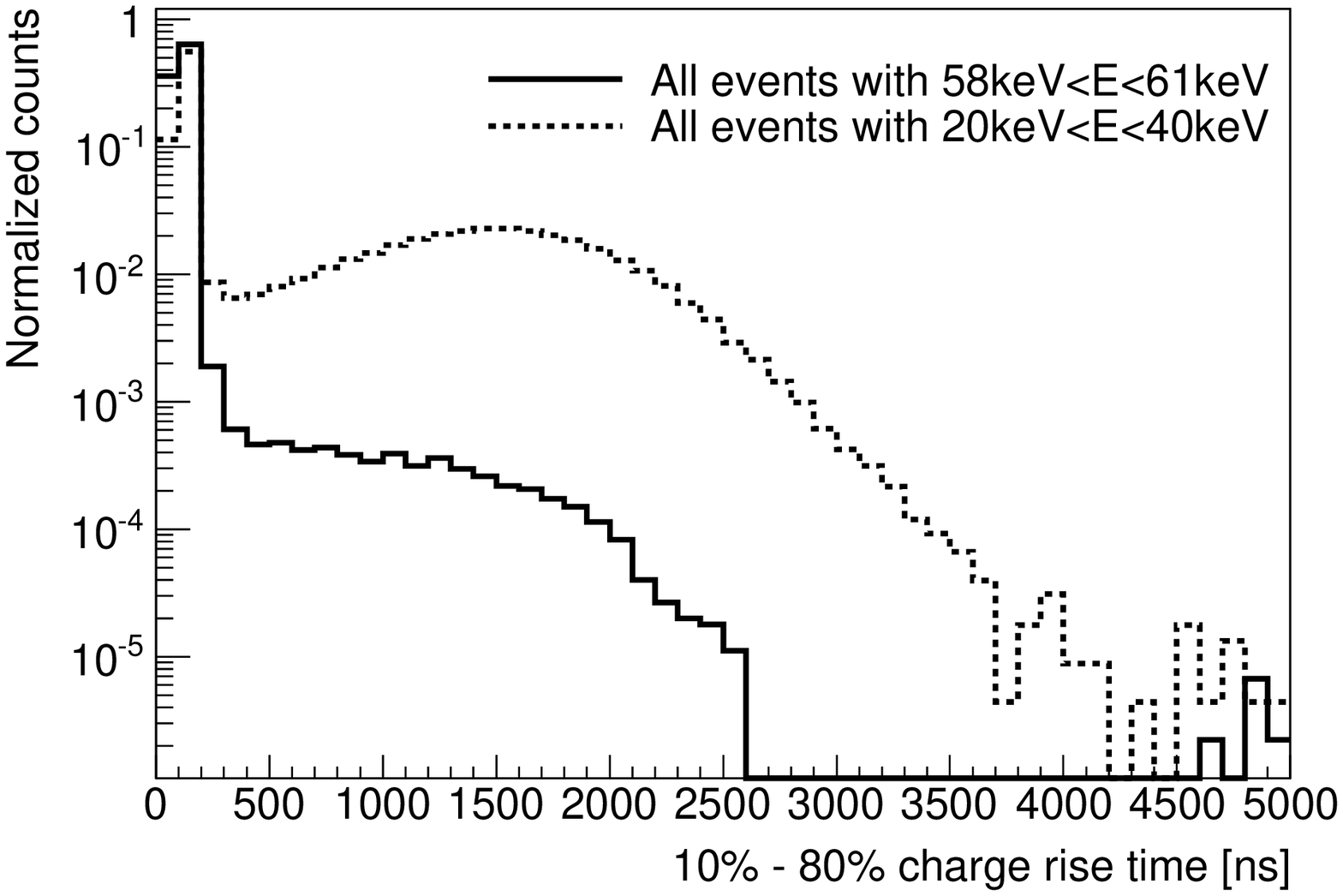} 
\caption{\label{fig:RiseTimeDistributionAtDifferentEnergies}Normalized distribution of the 10\% to 80\% rise times of the charge waveforms for events with energies in the 59.5\,keV photopeak (solid) and within the continuum between 20\,keV and 40\,keV (dashed). Interactions in the lithium-diffused layer result in a population of characteristically slow pulses contaminating the lower part of the energy spectrum.}

\end{figure}

 Fig. \ref{fig:RiseTimeDistributionAtDifferentEnergies} shows a comparison of the normalized 10\% to 80\% charge waveform rise time distributions of events with energies between 58\,keV and 61\,keV (near the 59.5\,keV photopeak) (solid line) and that of events in the continuum between 20\,keV and 40\,keV (dashed line). There is an evident increase in the number of slow pulses at low energies, highlighting how events near the lithium-diffused layer can contaminate the lower part of the energy spectrum.

\section{Determination of the transition layer depth}

In Section \ref{sec:lithickness}, it was shown that different profiles for the depth of interactions in a detector could be sampled by scanning the side of the detector. The data from the scan of the side of the detector are used here to characterize the depth at which slow pulses arise. This section describes a fit of the fraction of slow pulses as a function of the collimator $y$-position to distributions obtained from the Monte Carlo simulation discussed in Section \ref{sec:lithickness}. 

\begin{figure}[!htbp]
\centering
\includegraphics[width=0.8\textwidth]{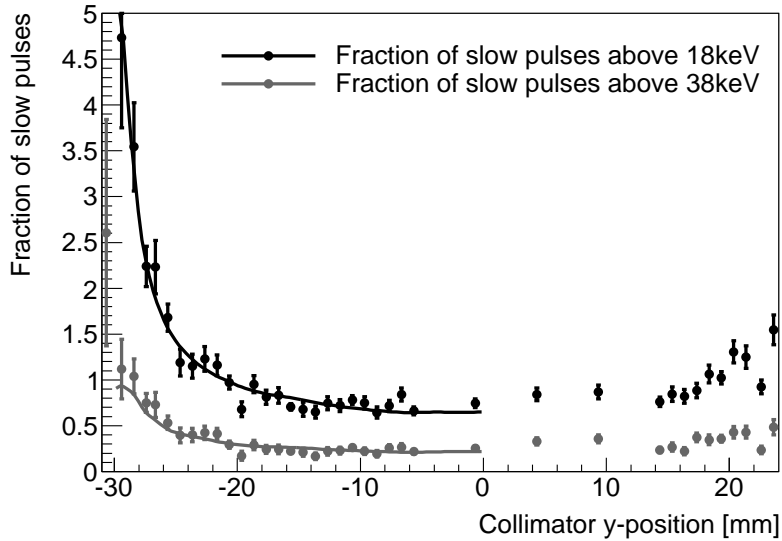}
\caption{\label{fig:TransitionFitComp} The fraction of SED events in the scan of the side of the detector as a function of the $y$-position of the collimator (in the same geometry as in Fig. \ref{fig:geometry_yz_full}). The fraction of slow pulses in the data is taken as the number of slow events with measured energies between $E_{min}$ and 58\,keV divided by the number of events in the 59.5\,keV photopeak. The distribution of slow events is shown for values of $E_{min}$=18\,keV and $E_{min}$=38\,keV. The data with the lower energy threshold have a larger fraction of SED events. These are fit to PDFs obtained from the Monte Carlo simulation (solid lines) to determine the depth at which slow pulses with an energy above $E_{min}$ are detected. It is worth noting that the Monte Carlo PDFs did not need to be normalized to the data and that the depth of the transition layer can be determined at any position from the ratio of slow pulses to full energy events independently from the source strength.}
\end{figure}

It was found that, in the LBNL PPC detector, a slightly cleaner sample of the slow pulses could be obtained using the asymmetry of the current pulse (see Fig. \ref{fig:PulseComparison}) rather than the 10\% to 80\% charge pulse rise time. A cut to select slow pulses in the data was derived from the ratio of the 10\% to 80\% current rise time to the 80\% to 10\% current fall time. The fraction of slow pulses for each position of the collimator was defined as the number of slow pulses with energy between $E_{min}$ and 58\,keV divided by the number of full energy events in the 59.5\,keV photopeak (determined in the same way as for the measurement of FCCD). The number of slow pulses between $E_{min}$ and 58\,keV was corrected by subtracting an energy-independent background of slow pulses estimated by integrating the number of slow pulses in a range from 61\,keV to 80\,keV. It was verified that the inferred statistical uncertainty from the background subtraction was consistent with data taken with no source present. The fraction of slow pulses as a function of the collimator position is shown in Fig. \ref{fig:TransitionFitComp} for $E_{min}$=18\,keV and $E_{min}$=38\,keV. As the the threshold for the minimum energy is lowered, the fraction of slow pulses increases, as expected from Fig \ref{fig:EnergyVsRiseTime_XY}.

The distributions of slow pulses as a function of the collimator $y$-position for different values of $E_{min}$ in the data were then fit to a set of PDFs derived from the Monte Carlo. The PDFs were obtained in a way similar to those that were used to determine the full charge collection depth. In this case, for each position of the collimator in the Monte Carlo data, the fraction of slow pulses is taken as the number of events that deposit their energy between the transition layer depth, TLD,  and FCCD divided by the number of events that deposit their full energy at a depth greater than FCCD (see Fig. \ref{fig:MCPDFS}). TLD($E_{min}$) is identified as the depth at which full energy depositions are detected with an energy above $E_{min}$. The denominator is simply the number of full energy events that was used previously to determine FCCD. The fraction of slow pulses in the Monte Carlo data corresponds to the ratio of the area in region 2 to the area in region 3 in Fig. \ref{fig:MCPDFS}.

A set of Monte Carlo PDFs were built with different values of the TLD. A function was then constructed to interpolate between these PDFs and fit the data in Fig. \ref{fig:TransitionFitComp} to determine the value of the transition layer depth. The position of the crystal and uncertainty from parallax were determined from the fits for the full charge collection depth in Fig. \ref{fig:LiThicknessFit} and allowed to vary within one standard deviation of those results. The fits were performed using different values of $E_{min}$ in the data. The solid lines in Fig. \ref{fig:TransitionFitComp} show the results of this fit to the data with $E_{min}$=18\,keV and $E_{min}$=38\,keV. Data were only fit for negative values of the $y$-position of the collimator to remove any effect that may be due to the copper rods in the mount. It was verified that the determined value of TLD was not affected when the fit range was extended, although the quality of the fit worsened.

\begin{figure}[!htbp]
\centering
\includegraphics[width=0.8\textwidth]{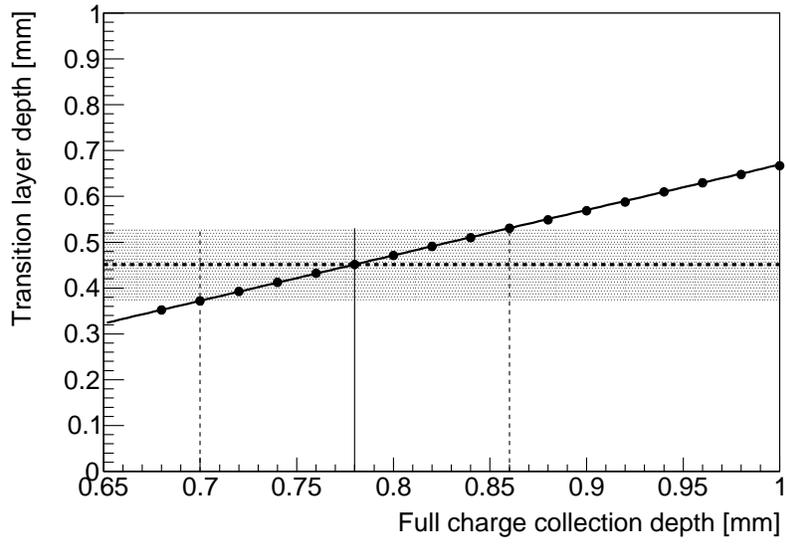}
\caption{\label{fig:transitionvsfulldepth}Transition layer depth (TLD) for slow pulses with $E_{min}$=26\,keV determined from a fit similar to those in Fig. \ref{fig:TransitionFitComp} as a function of the full charge collection depth (FCCD) that was used to generate the PDFs. The horizontal shaded area shows the uncertainty in TLD that is determined from the allowed range in FCCD obtained from the fit in Fig. \ref{fig:LiThicknessFit} (shown with the vertical lines). This is the dominant systematic uncertainty in the transition layer depth. FCCD was determined to be 0.78\,mm $\pm$ 0.08\,mm in Section \ref{sec:lithickness} and the resulting value for TLD is thus 0.48\,mm $\pm$ 0.08\,mm, for slow events with an energy above 26\,keV.}
\end{figure}

The dominant systematic uncertainty in determining TLD was the full charge collection depth. The two quantities are directly correlated as can be seen in Fig. \ref{fig:transitionvsfulldepth}, which shows the value of TLD that was determined for data with $E_{min}$=26\,keV as a function of FCCD. The statistical and systematic errors from the fit are are too small to see in Fig. \ref{fig:transitionvsfulldepth} and are comparatively negligible. The uncertainty on TLD was determined by using the range corresponding to FCCD = 0.78\,mm $\pm$ 0.08\,mm, as determined in Section \ref{sec:lithickness} and is illustrated by the shaded region in Fig. \ref{fig:transitionvsfulldepth}. The depth at which slow pulses with a minimum detected energy of 26\,keV arise was determined to be 0.48\,mm $\pm$ 0.08\,mm. The location where these slow pulses arise corresponds to region 2 in Fig. \ref{fig:MCPDFS}.

\begin{figure}[!htbp]
\centering
\includegraphics[width=0.8\textwidth]{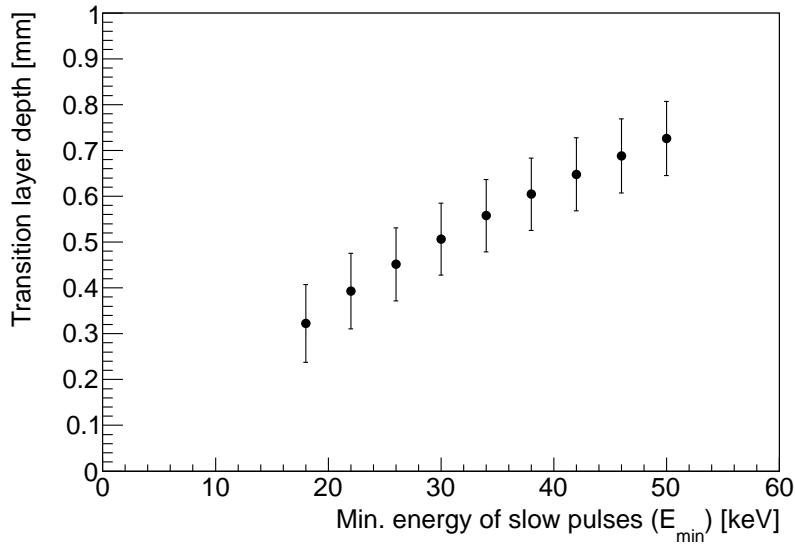}
\caption{\label{fig:minenergyvstransitiondepth} Transition layer depth as the minimum energy of the SED pulses is varied. As lower energy SED events are considered, the transition layer depth decreases. The truly dead layer of the detector, where no energy depositions would be detected, may be quite small, as the data are extrapolated to lower energies.  Understanding this curve for a detector is critical in estimating the background from slow pulses that one would observe at a very low energy threshold.}
\end{figure}

The analysis was performed for different values of $E_{min}$ ranging from 18\,keV (the energy threshold that was used to collect the data) to 54\,keV in 4\,keV increments. The value of TLD that was determined as a function of the minimum energy of the slow pulses is shown in Fig. \ref{fig:minenergyvstransitiondepth}. The detector, although capable of a much lower energy threshold, was only operated down to 18\,keV because of an instability in the preamplifier and the trend is not demonstrated to lower energies. Nonetheless, it is clear that high energy events that deposit their energy in the transition region result in slow pulses with a degraded measured energy. As the depth of an energy deposition increases, more energy is detected. Fig. 2 in Ref. \cite{Aalseth:2010vx} as well as data taken by the the {\sc Majorana} collaboration with the detector described in Ref. \cite{Aalseth2011692} show that the slow, energy-degraded, events from a 59.5\,keV $^{241}$Am gamma ray source extend to below 1\,keV. This is of significant concern to experiments operating with a very low energy threshold. The shape of the curve in Fig. 12 at low energies is unknown, and is likely different for different detectors, depending on how the n+ contact is fabricated

In order to understand this result, a similar set of analyses were performed using the detector described in Ref. \cite{Aalseth2011692}. This is a modified ``Broad Energy Germanium'' (BEGe) detector fabricated by CANBERRA and is similar to the one used by the CoGeNT collaboration. The modifications include a thicker lithium-diffused layer, a smaller point contact (4\,mm diameter) compared to the commercially available BEGe detectors, and a low background copper cryostat. The detector is operated as part of the {\sc Majorana} Low-background BEGe at the Kimbalton Underground Research Facility (MALBEK) experiment.

\begin{figure}[!htbp]
\centering
\includegraphics[width=0.8\textwidth]{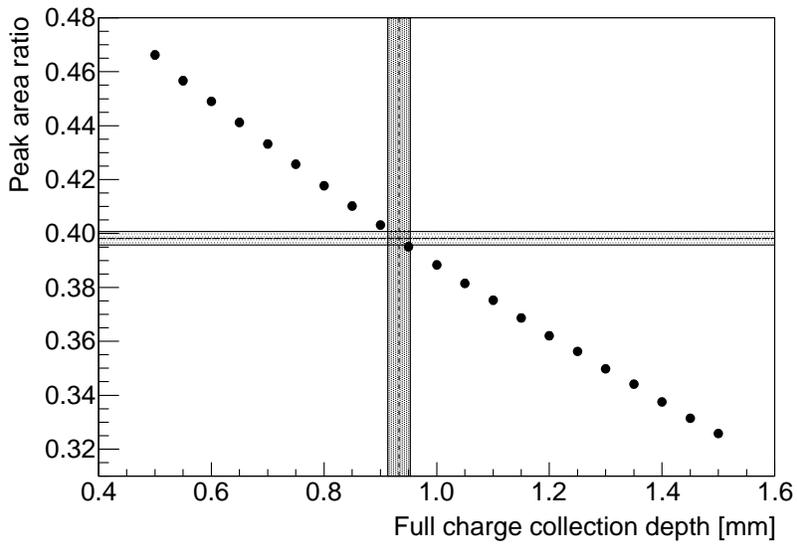}
\caption{\label{fig:baDeadLayerMeasurement_unbinned_0tail_1step} Determination of the full charge collection depth for the MALBEK detector using data from a $^{133}$Ba calibration source. The points with statistical error bars show the ratio of the number of counts in the 81\,keV and 356\,keV photopeaks from a simulation of the detector as a function of FCCD. The horizontal lines show the ratio that was measured in the data and the vertical lines show the value of FCCD determined by comparing the data and the simulation. For this detector, FCCD was determined to be 0.933\,mm $\pm$ 0.018\,mm (stat.) $\pm$ 0.092 (sys.).}
\end{figure}

The full charge collection depth for the MALBEK detector was obtained with a method similar to that described in Ref. \cite{Agostini:2010ke} using a $^{133}$Ba source. An energy spectrum was collected using a 0.9\,$\mu$Ci $^{133}$Ba calibration source placed 25\,cm from the end-cap of the detector (side opposite of the point contact). The ratio of the number of events in the 81\,keV photopeak to that in the 356\,keV photopeak was then determined. The experimental setup and cryostat were simulated using the Geant4-based MaGe simulation software \cite{Boswell:2010mr}. A full charge collection depth was implemented in the post-processing of the simulation and was adjusted so that the ratio of the number of counts in the two photopeaks matched the one measured in the data. Fig. \ref{fig:baDeadLayerMeasurement_unbinned_0tail_1step} shows the ratio from the simulation as a function of FCCD (points with statistical error bars) as well as the ratio obtained from the data (horizontal lines). FCCD for this detector (vertical lines in Fig. \ref{fig:baDeadLayerMeasurement_unbinned_0tail_1step}) was determined to be 0.933\,mm $\pm$ 0.018\,mm (stat.) $\pm$ 0.092 (sys.). In determining the systematic error, uncertainties in the length of the crystal and the thickness of the copper end-cap in the cryostat dominated over other effects such as uncertainties in the relative intensities of the photopeaks, live time, and the determination of the number of counts in the peaks.

\begin{figure}[!htbp]
\centering
\includegraphics[width=0.8\textwidth]{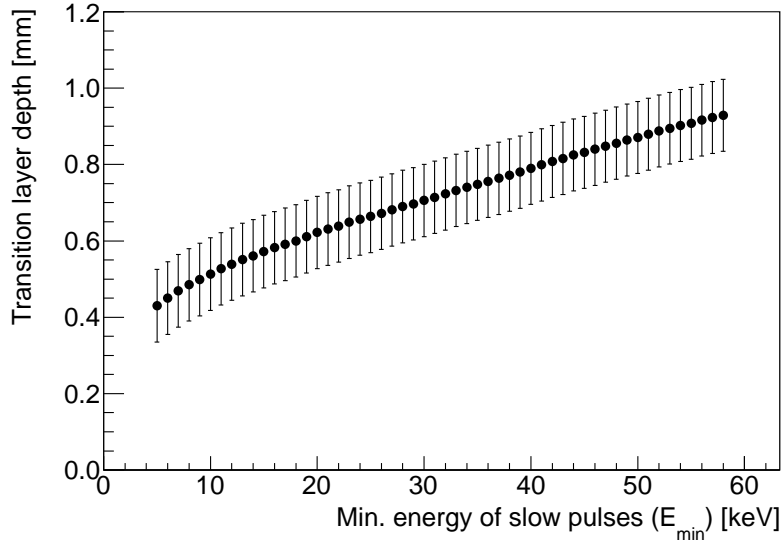}
\caption{\label{fig:transition_depth_MALBEK} Transition layer depth as the minimum energy of the SED pulses is varied in the MALBEK detector. Data  were recorded down to 5\,keV and show that this curve has a smooth behavior down to these low energies. The dominant systematic error shown on these data points is from the uncertainty in determining FCCD.}
\end{figure}

In order to probe the transition layer in the MALBEK detector, the end-cap of the detector was illuminated with an un-collimated 10\,$\mu$Ci $^{241}$Am source of 59.5\,keV gamma rays. The experimental setup was simulated and the transition layer depth was modeled in the same way as for the LBNL detector. SED pulses were selected based on the 10\%-90\% rising time of the charge waveform and backgrounds were determined using data taken with no source present. Fig. \ref{fig:transition_depth_MALBEK} shows the transition layer depth determined for the MALBEK detector as the minimum energy of the SED pulses is varied down to 5\,keV. The dominant systematic uncertainty in determining TLD for this detector was also found to be the uncertainty in determining FCCD (see above), which resulted in an 0.094\,mm systematic uncertainty included in the error bars in  Fig. \ref{fig:transition_depth_MALBEK}. The shape of the curve in Fig. \ref{fig:transition_depth_MALBEK} is similar to that in Fig. \ref{fig:minenergyvstransitiondepth} and does not show any discontinuities over this energy range.  The derivative of the two curves are different and suggest that the widths of the transition region and the fully dead layer are different for the two detectors and likely depend on the process used to fabricate the lithium-diffused layer.

The results presented in this work show that one can characterize the depth of the transition region and that a straightforward Monte Carlo simulation of the depth of energy depositions can be used to estimate the level of contamination from slow pulses. Although a very detailed scan of the LBNL PPC detector was performed, the transition layer depth and the full charge collection depth can be determined with few measurements. For example, FCCD can be determined using standard photopeak efficiency measurements. A particularly useful technique is to use gamma rays emitted from a $^{133}$Ba source and to compare the ratio of events in the low energy phototopeaks (around 81\,keV) to the number of events in the dominant line at 356\,keV to obtain FCCD, as was done for the MALBEK detector. These data could also be used to determine TLD by measuring the ratio of the number of slow pulses that come from the lines at 81\,keV to the number of full energy events in the corresponding photopeaks. A flood source measurement with $^{241}$Am was used for MALBEK to determine the transition layer depth on the surface opposite to the point contact. It is worth noting that these measurements only require a detailed knowledge of the absorbers between the calibration source and the detector and that they are independent of the calibration source strength.

The level of detail required to fully characterize a detector and quantitatively predict the number of SED pulses depends on an understanding of the uniformity of the lithium-diffusion. For the LBNL PPC in this work, the measured distributions of full energy events (Fig. \ref{fig:LiThicknessFit}) and of SED pulses (Fig. \ref{fig:TransitionFitComp}) could be fit over the entire range of $y$-positions of the collimator suggesting that the lithium-diffused layer along that axis is uniform. Although FCCD and TLD can be determined by a measurement at a single position on the detector, experiments requiring a thorough understanding of the SED contributions to their spectrum should perform a measurement to quantify the uniformity of the lithium-diffused contact. This can be achieved with a combination of flood source measurements, as was done for the MALBEK detector, and measurements performed using a collimated source.

\section{Conclusion}
 This paper presented data taken with the MALBEK detector as well as with a custom PPC detector fabricated at Lawrence Berkeley National Laboratory (LBNL). The mount holding the LBNL crystal and the geometry of the n+ contact of this system allowed effects from low energy gamma rays interacting in the lithium-diffused outer n+ contact of the detector to be examined in detail.

 The existence of a transition layer between the active region of the detector and the outer n+ lithium-diffused contact was confirmed. Charge depositions in this transition layer were observed to result in partial charge collection (energy degradation) as well as characteristically slow rise times of the corresponding charge pulses. A method was introduced to directly measure the full charge collection depth as well as the depth at which the transition layer begins. It was shown that energy depositions near the surface of the detector contacts can still result in detectable amounts of energy and that the truly dead layer can be smaller than is typically assumed. Data taken with the MALBEK detector suggested its transition and fully dead layers had different widths than those of the LBNL detector, indicating that the process for fabricating the lithium-diffused contact may play an important role.  Detailed spectral simulations of germanium detectors with thick lithium-diffused contacts need to account for these effects in order to reproduce spectral features at energies below known photopeaks.

 The slow, energy-degraded, (SED) pulses are of particular concern to experiments using germanium detectors with a very low energy threshold, such as for dark matter detection. Matters are complicated by the fact that cuts to remove these events based on pulse shapes (such as those used in Ref. \cite{Aalseth:2010vx}) become difficult to implement at energies of a few keV, where pulses are noisy and calculations must rely on effective de-noising techniques. Experiments using data from the lowest energies must demonstrate that contributions from SED events are understood. This is particularly important if there are gamma or x-ray lines whose SED events contribute to the continuum and a detailed knowledge of the spectrum is required. This work presented a method to characterize energy depositions in the lithium-diffused contacts which can be used to quantify the contributions to backgrounds from SED pulses.

 Mitigation strategies for experiments looking for signals at lower energies should include detailed characterizations of the detectors and of cuts based on pulse shapes, along with the demonstration that the associated systematic uncertainties are well understood. Detector development work and research into fabricating uniform and/or thinner hole-blocking contacts on the detectors, which could result in a sharp transition layer, are another research avenue. Promising work has been done to show that one can make thin, effective, hole-blocking contacts using amorphous silicon or germanium \cite{amman:2007} and yttrium \cite{hull:2010}. Furthermore, n-type point contact detectors \cite{luke_89} may have similar performance as p-type detectors at low energies; these can be a candidate for low energy experiments as they would almost entirely remove this source of background. However, if detectors with thin contacts are used, the benefit of shielding low energy backgrounds with the thick traditional contact of p-type detectors would be lost.

\bibliography{MJReferences}

\bibliographystyle{model1a-num-names}

\end{document}